# Hot-electron dynamics in plasmonic nanostructures

Jacob Khurgin[1], Anton Yu. Bykov[2], Anatoly V. Zayats[2]

[1]Electrical and Computer Engineering Department, Johns Hopkins University, Baltimore, MD, USA

[2]Department of Physics and London Centre for Nanotechnology, King's College London, Strand, London, WC2R 2LS, United Kingdom

**The coherent oscillations of mobile charge carriers near the surface of good conductors—surface plasmons—are been exploited in many applications in information technologies, clean energy, high-density data storage, photovoltaics, chemistry, biology, medicine and security. Light can be coupled to surface plasmons and trapped near the interface between a metal and an adjacent material. This leads to the nanoscale confinement of light, impossible by any other means, and a related electromagnetic field enhancement. Microscopic electron dynamic effects associated with surface plasmons are capable of significantly influencing physical and chemical processes near a conductor surface, not only as a result of the high electric fields, but also via the excitation of energetic charge carriers: holes below Fermi level or electrons above it. When remaining inside plasmonic media, these so-called hot carriers result in nonlinear, Kerr-type, optical effects important for controlling light with light. They can also transfer into the surroundings of the nanostructures, resulting in photocurrent, or they can interact with adjacent molecules and materials, inducing photochemical transformations. Understanding the dynamics of hot carriers and related effects in plasmonic nanostructures is essential for the development of ultrafast detectors and nonlinear optical components, broadband photocatalysis, enhanced nanoscale optoelectronic devices, nanoscale and ultrafast temperature control, and other technologies of tomorrow. This review will discuss the basics of plasmonically-engendered hot electrons, theoretical descriptions and experimental methods to study them, and describe prototypical processes and examples of the most promising applications of hot-electron processes at the metal interfaces.**

Nonequilibrium charge carriers are important in many fields of physics and chemistry and explored in metallic and semiconducting materials to control nonlinear optical response, photodetection, electronic tunnelling devices and chemical reactions, to name but a few[1-5]. Upon absorption of light in material, electrons and/or holes with excess energy (hot carriers) can be created. The excess energy depends on photon energy and a material band structure. In semiconductors, the photon energy should exceed a band gap. In contrast, in metals (or doped semiconductors), photon absorption can take place through both interband and intraband—within the conduction band—transitions. The latter is facilitated by plasmonic excitations.

The field of plasmonics has experienced steady progress through the development of our understanding of the optical properties of complex nanoscale metal structures to their application in sensing, information processing and nanomedicine[6-8]. Most of these applications are based simply on the strong field localisation and high field enhancement near a metal surface, found in plasmonic nanostructures. However, the phenomena associated with the effect of the high fields on the behaviour of electrons within the metal nanostructures have been largely ignored until recently. These effects arise from the same coherent oscillations of free electrons (surface plasmons). They are capable of influencing physical and chemical processes near the metal interface, not only because of the associated enhanced electric fields but also as a result of the modification of the electron energy distribution and/or transfer of energetic electrons from the metal to adjacent molecules or materials in the surroundings.

Plasmonic excitations, such as surface plasmon polaritons (SPPs) on smooth metal interfaces or localised surface plasmons (LSPs) on nanostructures and nanoparticles, provide strong light absorption without the need for a band-gap and efficient hot-carrier generation. The created carrier distribution is initially non-thermal and evolves in complex ways through internal thermalisation as well as external processes, such as carrier or energy transfer into the surrounding environment, influencing a variety of processes. The hot-carrier relaxation time is determined by the interplay between electron-electron, electron-phonon scattering and carrier extraction probabilities and can be controlled by engineering the nanostructure shape and pathways of interaction with the surrounding environment.

The generation of hot electrons in plasmonic structures via surface plasmon excitations has several significant advantages: 1) high generation efficiency related to strong light absorption via plasmonic resonances; 2) the broadband nature of the hot-electrons related to the broadband nature of plasmonic excitations (not limited by semiconductor bandgaps, their energy can be tuned simply by changing the wavelength); 3) the high density of hot electrons in a spatially well-defined volume near the plasmonic surface, from where they can be efficiently extracted before they thermalise in the metal by engineering resonant charge transfer to proximal molecules or semiconductors.

In thermal equilibrium, the energy distribution of the electrons is determined by a temperature which is the same as a lattice temperature. Nonequilibrium hot carriers have a distribution function which cannot be assigned a single temperature (nonthermal distribution). The thermalisation process results in the thermal distribution of the hot carriers, which is characterised by a temperature which is distinct from lattice temperature, and the lattice (phonon) temperature can differ from the temperature of the surroundings. Thermally distributed hot carriers, which are on average much less energetic than non-thermal ones, will then further lose the energy to phonons, heating the lattice. These hot-electron thermalisation processes take place on different time scales, and hot electrons with different energies are important in the context of diverse physical and chemical processes.

In this Review, we shall discuss a microscopic picture of the hot-electron generation and evolution in plasmonic nanostructures and related macroscopic effects, including control of nonequilibrium electrons in nanostructures as well as plasmonic nanostructures as a source of hot electrons in semiconductors and molecules. We start with a description of the light absorption in plasmonic nanoparticles, introduce the relevant thermalisation processes and discuss the hot-carrier generation rate. Both CW, such as solar, and pulsed, such as femtosecond laser, excitations are considered and compared. After comparing the hot-carrier generation mechanisms, the injection of hot electrons across the metal interface is discussed, particularly at the metal-semiconductor interfaces. The chemical effects related to hot-carrier injection into surroundings are also considered, and their influence on hot-carrier dynamics is overviewed. Finally, nonlinear optical response induced in plasmonic nanostructures by hot-carrier excitation is discussed, emphasising their role as a tool for understanding hot-electron processes.

## Hot-electron generation and relaxation.

### Discrete (quantum) nature of hot carrier generation and decay.

To better understand the importance of the quantum character of the processes that take place when hot carriers are generated and decay, from the very start, we shall point to the difference between laser illumination, which allows one to achieve power densities of GW/cm$^2$ and the illumination with thermal and other incoherent light sources limited to 100s of W/cm$^2$ irradiance (e.g., 100 W/cm$^2$ is roughly 1000 times that of the irradiance of the sun on the equator). It is the latter, CW excitation case with low to moderate intensities, that is beneficial for practical applications in solar-driven chemistry, while the former is important for high-value chemical synthesis as well as serves as an excellent tool for the study and understanding of carrier dynamics on the femtosecond scale and applications in nonlinear optics  Let us consider an ensemble of identical plasmonic spherical nanoparticles of diameter d = 5-30 nm illuminated with light at the wavelength corresponding to the LSP resonance (Fig. 1a).

Under CW irradiation (e.g., $I_{in} = 100 \, W/cm^2$), a fraction of nanoparticles excited at any given time can be estimated by noting that the electric field in the LSP mode gets enhanced approximately a factor of $Q = \omega/\gamma$, where $\gamma$ is the total (radiative and nonradiative) decay rate of the LSP mode[9,10]. This factor is $Q \sim 10-20$ for good plasmonic metals, such as Au or Ag,[11] and even less than that for small nanoparticles where the Landau damping is prevalent.[12] Therefore, the energy density in the LSP mode can be found approximately as $Q^2 I_{in}/c$, where $c$ is the speed of light, and the total energy residing on a given nanoparticle at a given time is $U_{LSP} \sim Q^2 (I_{in}/c) V_{eff}$, where $V_{eff}$ is the effective volume of the LSP mode that is commensurate with the volume of the nanoparticle itself.  The number of LSPs per one nanoparticle (which is the same as the fraction of all nanoparticles that have an LSP residing on them at a given time) is, therefore, $N_{LSP} \sim Q^2 (I_{in}/c) V_{eff}/\hbar\omega$. For the illuminating light with $\hbar\omega = 2 \, eV$ and $Q = 20$ (corresponding to $\gamma \sim 1.5 \times 10^{14} s^{-1}$), this estimate gives $N_{LSP} \ll 1$, ranging from $2 \times 10^{-7}$ to $6 \times 10^{-5}$ depending on the nanoparticle size (Fig. 1b). This means that, for example, for a nanoparticle with a diameter of 20 nm and CW illumination, at a given time only 0.002% of the nanoparticles actually have a LSP on them while the vast majority do not have any. This situation is depicted in Fig. 1a, where only a few nanoparticles are currently excited, i.e., have an SPP on them (shown in yellow) or carry hot carriers generated as a result of the LSP decay (shown in red), while most of the nanoparticles remain "cold", meaning that the hot carriers that had been previously excited in them have already thermalised and then transferred their energy to the lattice.

Even in more sophisticated plasmonic arrangements involving various nanoantenna or nanofocusing geometries, the total field enhancement does not exceed $Q^2$ [13-16]. It is easy to see that even for nanoparticles as large as 50 nm the condition $N_{SPP} \ll 1$ is still maintained. It should take $I_{in} > 5MW/cm^2$ to achieve a situation where each nanoparticle has on average more than a single LSP – this situation can, of course, occur for pulsed laser excitation[17-20].

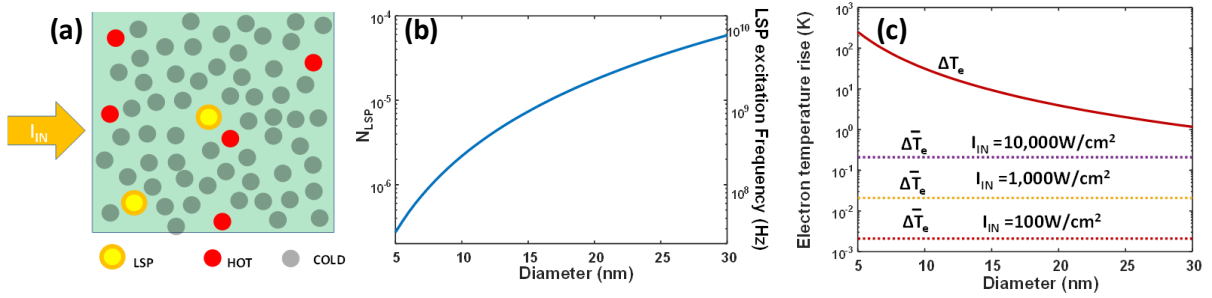

**Fig. 1.** Discrete character of hot carrier excitation in nanoparticles and its implications. (a) Excitation and decay of LSPs and nonequilibrium carriers in the nanosphere array illuminated by a CW light with irradiance $I_{in}$ at a given moment in time. LSPs (orange rings) are excited on a few nanoparticles shown as yellow circles, nonequilibrium hot carriers are excited in a few "hot" nanoparticles tinted red, while the electrons in the vast majority of nanoparticles remain "cold" i.e. close to the lattice temperature $T_L$ (although $T_L$ may itself be significantly elevated relative to the ambient temperature). (b) Probability of a LSP being excited on a given nanoparticle at a given moment in time $N_{LSP}$ and the frequency of excitation $f_{SPP}$ for the input irradiance $I_{in} = 100W/cm^2$. (c) Instant rise of the electron temperature following the decay of the LSP $\Delta T_e$ (solid line), which is independent of $I_{in}$, and the rise of the time-averaged electron temperature relative to the lattice temperature $\Delta \bar{T}_e$ (dashed lines) if one neglects the quantum nature of the absorption process for three different values of $I_{in}$.

Even if the concentration of the nanoparticles is high enough to achieve very strong (even total) absorption of the incident light, one can easily estimate the frequency with which the LSPs are generated on a given nanoparticle as simply $f_{LSP} = \gamma N_{LSP} = Q(I_{in}/\hbar c)V_{eff}$ (Fig. 1b). For the aforementioned case of $d=20$ nm, $f_{LSP} \sim 2.5\ GHz$. Therefore, the LSP on an individual nanoparticle is excited every 400 ps and then decays within brief time $\tau_{spp} \sim \gamma^{-1} \sim 7\ fs$ [17,19,21,22]. Since the radiative decay rate of the LSP is on the order of $\gamma_{rad} \sim \omega V/\lambda^3 \sim 10^{11} s^{-1}$ [23], it can be neglected and $\gamma$ can be considered to be almost entirely nonradiative, i.e., each time LSP decays, an electron-hole pair gets generated in the metal.

The generated electron-hole pair in turn survives for a very short time, as it is subject to both electron-electron (EE) and electron-phonon (EP) scattering processes [24] with scattering rates $\gamma_{ee}$ and $\gamma_{ep}$ respectively. While these rates in noble metals are roughly comparable (on the scale of $10^{13}$-$10^{14}$ s$^{-1}$) [25], the actual rate of energy transfer to the lattice, $\gamma_{EL}$ is significantly lower than that because the energy of a typical phonon (i.e., the Debye energy) is many times smaller than LSP energy $\hbar\omega$. Thus it takes many EP scattering events to transfer all the energy from the carriers to the lattice and to subsequently raise its temperature by $\Delta T_L = \hbar\omega/c_L N_e V$ per each decayed LSP, where $c_L$ is molar specific heat of the lattice and $N_e \sim 6 \times 10^{22} cm^{-3}$ in metals.[26] At the same time, since in each EE scattering event the energy of a hot carrier is shared between three carriers (the original electron(hole) is scattered and a new electron-hole pair is created when the electron is promoted from below to above Fermi level), it takes only a very few of those events to spread the initial energy $\hbar\omega$ between all the electrons near the Fermi level and raise the electronic temperature by $\Delta T_e = \hbar\omega / c_E N_e V$, where $c_E \ll c_L$ is specific heat of the electrons[24]. Therefore, one can introduce the electron thermalization time $\tau_E$ which is only a few times longer than EE scattering time $\tau_{ee} = \gamma_{ee}^{-1}$ and is many times shorter than the electron cooling rate $\tau_{EL} = \gamma_{EL}^{-1}$, which is at least an order of magnitude longer than EP scattering time $\tau_{ep} = \gamma_{ep}^{-1}$.

Let us now consider the dynamics of the processes occurring in a given nanoparticle (Fig. 2):

(a) First, at time $t_0$, an LSP is generated at the "cold" nanoparticle (upper row), where the electrons are in equilibrium with the lattice (middle row) has the temperature $T_e = T_L$ (lower row).

(b) Then, at time $t_1 \sim t_0 + \tau_{LSP}$, LSP decays engendering a single "primary" or "first generation" electron-hole pair, creating a non-thermal carrier distribution which cannot be assigned an electron temperature.

(c) Roughly at time $t_2 \sim t_1 + \tau_{ee}$, primary electron (hole) carriers undergo a collision with the electron below the Fermi level and its energy is shared between three second generation carriers – two electrons and one hole (or two holes and one electron). The second-generation carriers follow the same routine and, with only a few generations at time $t_3 \sim t_2 + \tau_E$, a thermal quasi-equilibrium is reached. The term "quasi" is used here to indicate that the carrier distribution strictly cannot be described by a simple Fermi-Dirac function with a well-defined single temperature $T_e$ [27][28]. For a simple analysis, we can estimate the electron temperature rise using the electron specific heat $c_E = \frac{\pi^2 k_B^2 T_L}{2 E_F}$, where $E_F$ is the Fermi energy ($E_F = 5.5$ eV for both Au and Ag). The electron temperature rise $\Delta T_e = 2\hbar\omega E_F / \pi^2 k_B^2 T N_e V$ amounts to less than 4 K for a d=20 nm nanosphere and exceeds 10 K for d<15 nm and 100 K for d<7 nm (Fig. 1c). During this time period, the lattice temperature remains practically unchanged.

(d) Following that, at time $t_4 \sim t_2 + \tau_{EL}$, all the energy is transferred to the lattice whose specific heat $c_L = 3k_B$ is about $130$ times higher than the specific heat of the electron gas $c_E$, which leads to a very insignificant increase $\Delta T_L \sim 0.03 K$ over the average lattice temperature $\bar{T}_L$ which in turn can be found as $\bar{T}_L = T_0 + \Delta T_L f_{SPP} \tau_{LA}$, where $T_0$ is the ambient temperature and $\tau_{LA}$ is the lattice cooling (or lattice to ambient heat transfer) time, which is determined by the environment in which nanoparticle is placed and can be as long as milliseconds; so that $\bar{T}_L - T_0$ may approach 100 K and more. Such a strong temperature increase will definitely contribute to the hot carrier injection from metal into adjacent dielectric or semiconductor or to chemical processes on a metal surface.[29-32]

(e) By the time $t_4 \sim t_0 + f_{LSP}^{-1}$, i.e, just before the LSP is excited again both electron and lattice are once again at equilibrium with temperature $\bar{T}_L$ and the process repeats.

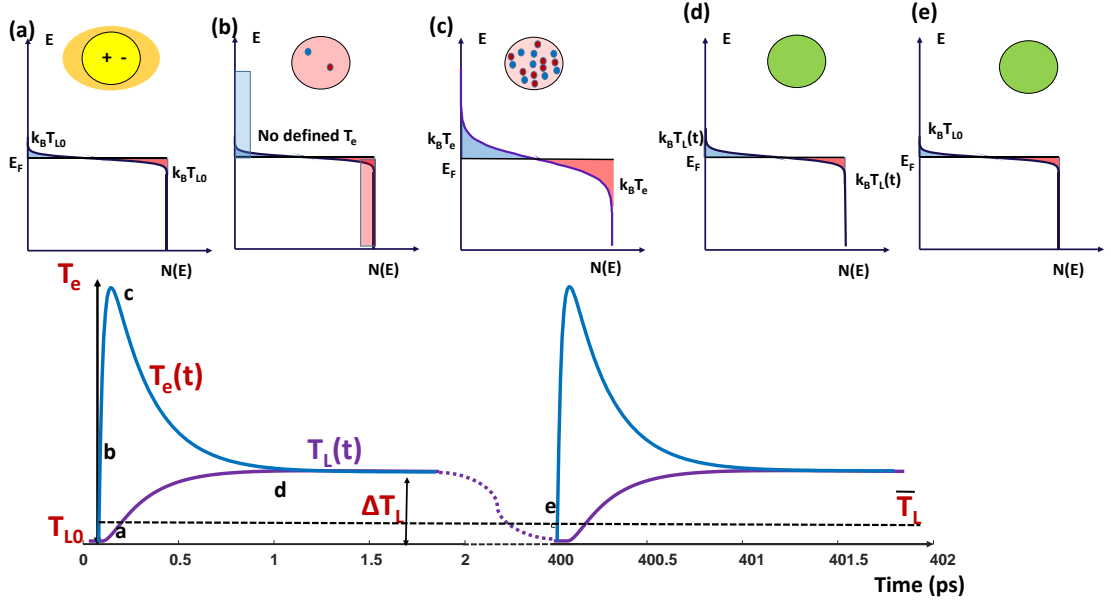

**Fig. 2**. Evolution of hot carriers in a plasmonic nanoparticle. (a-e) Energy distribution of electrons in a nanoparticle at different times (a) before and (b) immediately after SPP decay, (c) after thermalization of carriers and establishing an electron temperature, (d) after establishing thermal equilibrium with a lattice, (e) immediately before the next SPP gets excited. (f) Evolution of the electron and lattice temperatures (not to scale as $\Delta T_e \gg \Delta T_l$ ) through the stages (a)-(e).

The increase of an instant electron temperature $\Delta T_e$ does not depend on the incoming power density $I_{in}$ and is only a function of the nanoparticle volume. Increasing input power will only lead to the increase of the frequency with which the LSP is excited on a given nanoparticle (or, equivalently to the number of excited nanoparticles at any given time) but the temperature rise will remain the same. This, as well a simple fact that the (lattice) temperature of the nanoparticle is time dependent even though the illumination is CW, provide clear evidence of the discrete, quantum nature of the processes that take place. If one disregards the quantum nature of absorption, then under CW illumination, one simply obtains the average steady state rise of the electron temperature $\Delta \bar{T}_e = D \Delta T_e$ where the duty cycle $D = f_{SPP} \tau_{EL}$ so that, for $\tau_{EL} \sim 200 fs$, one obtains a much smaller temperature rise $\Delta \bar{T}_e \sim 0.001 \Delta T_e$ (Fig. 1c). In this case, even for very high input powers, the average rise of electron temperature is negligibly small [29,30,33].

To see the impact of this quantisation, let us consider thermionic emission of the electrons from a plasmonic nanoparticle to surroundings across the barrier Φ. In the absence of light this rate is $R_{th,0} \sim \exp(-\Phi / k\bar{T}_L)$. So, the relative increase of that rate due to average rise of electron temperature

$$\Delta R_{th}\left(\Delta \bar{T}_e\right)/R_{th,0} \approx \exp(\Phi \Delta \bar{T}_e / k_B \bar{T}_L^2) - 1 \qquad (1)$$

is negligibly small for tiny $\Delta \bar{T}_e$ as in Fig. 1c. On the other hand, if one properly follows the discrete nature of hot-carrier generation, one should use $\Delta T_e = \Delta \bar{T}_e / D$ in place of $\Delta \bar{T}_e$, but then multiply the rate by the duty cycle $D$, to obtain

$$\Delta \bar{R}_{th}\left(\Delta T_e\right)/R_{th,0} \approx D[\exp(\Phi \Delta \bar{T}_e / k_B D \bar{T}_L^2) - 1] \; . \qquad (2)$$

For $I_{in} = 100 \; W/cm^2$ $\Delta \bar{T}_e = 2 \times 10^{-3} K$ and for 20nm nanoparticle $D = 10^{-3}$ one obtains $\Delta \bar{R}_{th}\left(\Delta T_e\right)/\Delta R_{th}\left(\Delta \bar{T}_e\right) \approx 1.05$ but for 5nm nanoparticle $D = 2 \times 10^{-5}$ and $\Delta \bar{R}_{th}\left(\Delta T_e\right)/\Delta R_{th}\left(\Delta \bar{T}_e\right) \approx 30$! Nevertheless, the absolute value of the increase is still only $\Delta \bar{R}_{th}\left(\Delta T_e\right)/R_{th,0} \approx 0.003$. This "granular" character of absorption is preserved not just for LSPs on nanoparticles but also for surface plasmon polaritons (SPPs) propagating along the metal interface, important, for example, for the absorption in photodetectors employing metal/semiconductor interfaces. For a waveguide of length $L$ in which the SPPs with an average power $\bar{P}$ propagate with a group velocity $v_g$, an average number of SPP quanta present at any given time is $N_{SPP} = \bar{P}L / v_g \hbar \omega$. For $\bar{P} = 1 \mu W$ and $L = 10 \mu m$ it amounts to less than one SPP.

*What happens under femtosecond excitation?*

The situation is radically different in the case of ultrafast excitation when the pulse duration is comparable to the hot electron relaxation times, i.e., under femtosecond excitation. If one considers 100 fs pulses with 80 MHz repetition rate and an average power of 1 W focused in a 100 $\mu m$ spot, the peak intensities of $I_{in} \sim 10^9 W/cm^2$ are readily achievable and hundreds of LSPs can be simultaneously excited on a single nanoparticle. Since pulse length is shorter than $\tau_{EL}$ the entire energy of the absorbed pulse gets accumulated in the energy of hot carriers, hence the rise of electron temperature is hundreds or even thousand times larger than one caused by absorption of a single quantum of energy, its rise is obviously proportional to the excitation power, and the electron temperature rise of a few thousands K is not unreasonable. Remarkably, the dynamics of hot carriers and their temperature for femtosecond excitation would look very similar to Fig. 2 with magnitude of $\Delta T_e$ scaled up by a number of LSPs excited and the interval between pulses being regular and equal to the pulse repetition rate, rather than random. This situation is regularly encountered in nonlinear optical studies.

*Take home points.* When considering the generation and decay of surface plasmons in nanoparticles or thin films, one must always take into account the discrete character of all the processes taking place inside metal. The instant rise of the electron temperature may cause rise in thermionic emission, and, therefore, increased photo-injection and photocatalysis only for the smallest spherical nanoparticles under CW excitation. For large barriers, the absolute value of injection would still be very low. At the same time, if the lattice cooling time $\tau_{LA}$ is long (as it can be if the sample is thermally isolated from the environment), an average rise of the lattice temperature $T_L$ may be high enough to cause a conventional thermionic emission. *Therefore, in order to achieve a carrier injection over the barrier, the non-equilibrium carriers must be*

*present (before they thermalize to some average electron temperature), i.e. the primary electrons and holes that have not experienced a single EE scattering event.* It should be also noted that if the shape of the nanoparticle is different from the spherical or in the case of complex hereto-nanoparticles with different material components, the spatial distribution of the field enhancement is nonuniform and the hot-electron generated nonuniformly in the nanoparticle. In this case, the mode volume ($V_{eff}$) is not directly related to the size of the nanoparticle and the rate of the hot electron generation is proportional to the local absorption defined by the local field, which benefits sharp edges or junction in the hetero-nanoparticles. In the case of weakly absorbing nanostructures, the rate can be enhanced by engineering dark electromagnetic states to trap the excitation light and promote absorption and hot carrier generation. In the following sections, we describe how these carriers are generated and how they get injected from metal into semiconductor, dielectric or molecule.

**Four mechanisms of hot carrier generation in metals**

Let us now review the processes that lead to the decay of LSPs and SPPs and hot carrier generation. This issue has been widely investigated [34-44], yet some of the assumptions made in the previous works are ambiguous. In particular, it concerns the use of classical concepts to describe the LSP and SPP decay. Surface plasmons can be correctly described as collective oscillatory motion of the entire Fermi gas of the carriers with frequency $\omega$, but its decay cannot be represented as gradual loss of energy at each half-cycle of the oscillations to some "friction force" [34-39]. As discussed in previous section, the surface plasmon is a quantum object with a well-defined energy $\hbar\omega$ and it can only lose this energy in an instant process in which the LSP/SPP gets annihilated and at least two new (quasi)particles are created in accordance with energy and momentum conservation laws. It is important to emphasize that "friction" loss present in the classical Drude theory is a combination of electron-electron and electron-phonon (or electron-impurity) interactions and can be adequately described by the widely used in condensed matter physics second-order perturbation theory[24] (as shown below). LSP/SPP decay in a metal is similar to phonon-assisted photon absorption in the indirect bandgap semiconductor. The quantum nature of the LSP/SPP decay is manifested in the fact that while a classic Drude "friction" rate $\gamma$ behaves as $T^5$ [24] at temperatures below the Debye temperature, at optical frequencies, the LSP/SPP decay is still very fast [45,46] even at cryogenic temperatures because spontaneous emission of large wavevector phonons becomes allowed as long as a photon energy exceeds the Debye energy.

The other issue largely overlooked in the literature is the role of the electron-electron (EE) interaction assisted processes in the LSP/SPP decay. These interactions are indeed negligibly small at low frequencies but, as frequency increases, the contribution of the EE scattering grows as $(\hbar\omega)^2$ and, at optical frequencies, the contribution of the EE scattering becomes comparable to the scattering by phonons and defects as numerous experimental results confirm [25,47,48].

Yet another issue concerns the holes excited in the d-shells of noble metals. Their potential energy relative to the Fermi energy may be as large as the exciting photon energy $\hbar\omega$, their kinetic energy is, however, much smaller than that and the d-shell holes excited by the interband processes usually do not reach the surface of the metal and decay after only a few nanometers.

Last, but not least important misconception is that the LSP/SPP decay in metals is somehow different fundamentally from the photon absorption in dielectric or semiconductor. In fact, a photon in dielectric is also a polariton, formed by coupled oscillations of the electric field and collective oscillations of bound electrons in the valence band [49]. A significant fraction of the total energy is contained in the potential energy of these bound electrons. In the case of the surface plasmon polaritons, the only difference is that a large fraction of the energy is now contained in collective oscillations of free carriers (as their kinetic energy) [50]. Hence, the decay of SPP and photons is described by the same interaction of oscillating electric fields with the single electronic states in the medium (interaction Hamiltonian contains only electric field and the wavefunctions of initial and final single electron states). Collective electron excitations per se do not participate directly in the interactions and simply act as the "reservoirs" where the energy is stored during the half cycle when the electric field is small. This is equivalent to the photon-matter interactions where equal amounts of energy are carried by the electric and magnetic fields, but only electric field interacts directly with the matter in most cases. The electric field of LSPs/SPPs alone determines how they decay and what carriers are excited as a result of this decay. With this understanding, we will now consider the mechanisms that cause hot-carrier excitation one by one.

*Direct Interband Absorption.*

First mechanism is the direct interband (ib) absorption[41] between the inner (4d or 5d) and outer (5sp or 6sp) shells of noble metals (Fig. 3a). The energy gap separating the highest level in the d-shell and the Fermi level residing in the hybridized sp-shell $E_{ds}$ is close to 2 eV for Au and 3 eV for Ag. When direct absorption takes place, the kinetic energy of the electron generated in the s-band (relative to the Fermi level) is only $E_{ib} < \hbar\omega - E_{ds}$, as can be seen in the probability of the carrier energy distribution $F_{hot,ib}(E) = 1/(\hbar\omega - E_{ds})$. If one wants to consider hot carrier injection across the barrier on the order of 0.5-1eV, only short-wavelength (blue and UV) radiation can be used for this purpose with Au or Ag and majority of metals due to their band-gap energies. The angular distribution (with respect to the metal surface of the non-equilibrium carriers generated via interband absorption is uniform.[27]

The holes generated in d-shell have large potential energy relative to the Fermi level, but their kinetic energy is quite low, and, more important, the d-band is narrow and d-electron's velocity is at least an order of magnitude lower than Fermi velocity[51], the mean free path is also very short, less than a nm. The holes generated in the bulk of the metal decay long before they reach the interface. For this reason, the interband absorption, as it competes with other absorption mechanisms, only reduces the efficiency of hot-carrier extraction and if one wants to develop hot carrier applications in blue-UV spectral range, the logical step is using aluminium which does not have interband absorption (due to absence of d-shell) in that region.[52]

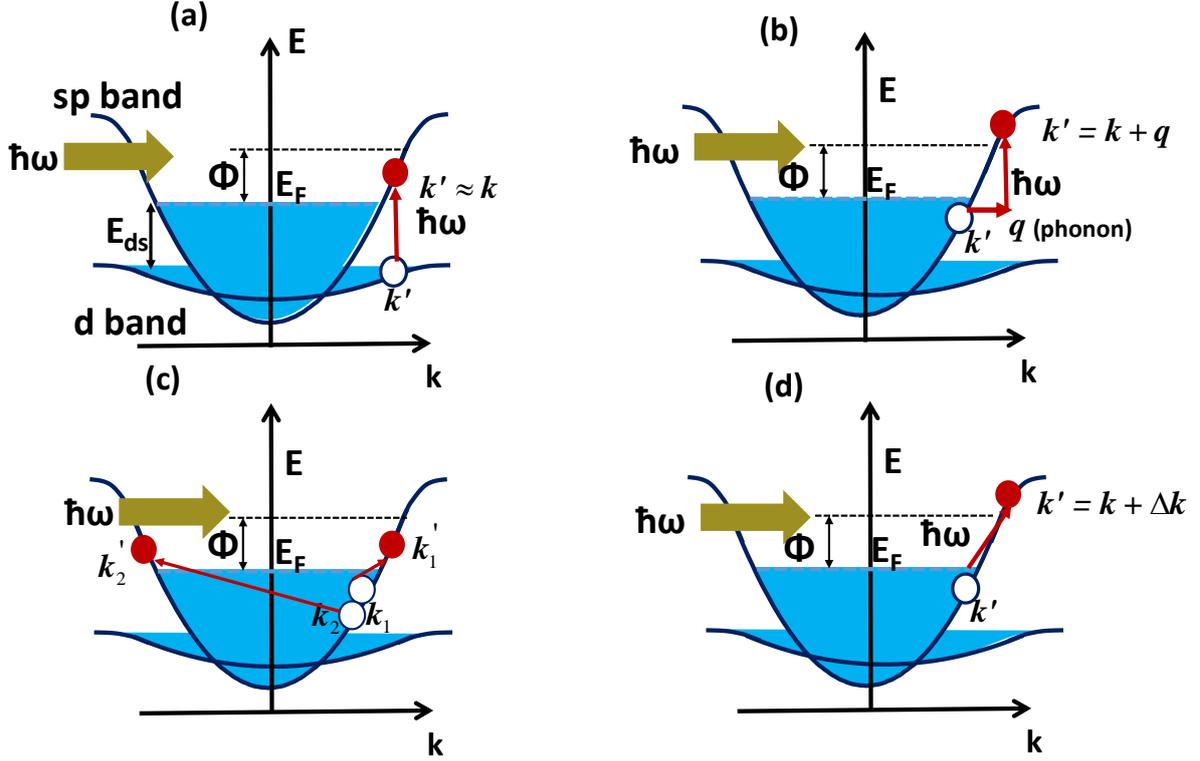

**Fig. 3.** Four mechanisms of electron hole pairs generation in metals. (a) Direct (momentum conserving) interband transition resulting in the low average kinetic energy of electrons and holes. (b) Phonon (or defect/impurity) assisted transition resulting in the average energy of electrons or holes being ℏω/2. (c) Electron-electron Umklapp scattering assisted transition with two electron-hole pairs generated for each LSP/SPP with the average kinetic energies being ℏω/4. (d) Landau damping or surface collision assisted "tilted" transition resulting in the average energy of electrons or holes being ℏω/2. The carriers generated via processes (b) and (d) are far more likely to have energy sufficient to overcome the surface (Schottky) barrier Φ and be injected into the adjacent semiconductor or dielectric or molecule.

*Phonon or defect assisted decay*

The remaining three LSP/SPP decay mechanisms are *intraband*, i.e. they involve absorption between two states with different electron momenta (wavevectors) in the same s-p-band. This momentum mismatch needs to be somehow compensated. The momentum conservation can be delivered by either a phonon or an impurity (defect) providing a wavevector $q$ (Fig. 3b). As the result, LSP/SPP is annihilated, a hot electron and hot hole, each with an average energy of $\hbar\omega/2$ are generated. The energy distribution of the "first generation" of hot carriers is $F_{hot,ph}(E) = 1/\hbar\omega$ with $E_F < E < E_F + \hbar\omega$ for hot electrons and $E_F > E > E_F - \hbar\omega$ for hot holes. Superficially, this process is similar to "Drude"-like absorption arising from the imaginary part of dielectric constant $\varepsilon_i = \gamma\omega_p^2/\omega^3$ and the SPP damping rate due to this process is $\gamma_{ph}(\omega) = \left\langle \tau_{ep}^{-1}(E) \right\rangle_E$, where the electron-phonon (or defect) scattering rate is averaged from $E_F - \hbar\omega$ to $E_F + \hbar\omega$ At the same time, in the Drude formula, the scattering rate $\gamma$ is evaluated near the Fermi level, because that is where both initial and final states of scattering carriers reside when the photon energy is small. At low temperatures, phonon scattering near the Fermi level behaves as $T^5$ [24] and the actual Ohmic resistance at low frequencies becomes significantly lower. However, when the photon (or LSP/SPP) energy $\hbar\omega$ substantially exceeds the Debye energy $\hbar\theta_D$ (which is few tens of meV), the phonon scattering rate exhibits much weaker temperature dependence [45] and, in fact, stays within the range of $\gamma_{ph} \approx 3 \times 10^{13} s^{-1}$ for Ag $\gamma_{ph} \approx 10^{14} s^{-1}$ for Au [53]. It is important to stress, once again, that the LSP/SPP decay is a quantum, not a classical process. There is no such thing as so-called "classical" [39], "resistive" [34,36] or "friction [37]" contribution to the SPP decay, in which a "sea" of multiple low energy carriers supposedly instantly created. The energy of the SPP is almost entirely (minus small phonon energy) transferred to just two hot particles–electron and hole– and is not dissipated to a bath of multiple carriers near the Fermi level.

Nevertheless, some classical analogies remain true even in the quantum picture. Since classically the carriers are accelerated along the direction of the optical field, one would expect the photoexcited hot electrons and holes to preferentially travel in that direction. Indeed, detailed calculations show that the photoexcited carriers have normalized angular distribution relative to the direction of field (which is often normal to the surface) is[41] $R_{ph}(\theta) = \frac{3}{4}\cos^2\theta + \frac{1}{4}$ The electric field near hot spots in many plasmonic nanostructures is close to being normal to the surface, therefore, the fraction of hot carriers going towards the surface is twice as large than for the uniform distribution in the case of interband and EE assisted scattering.

*Electron-electron scattering assisted decay*

The third mechanism of the SPP decay involves the electron-electron (EE) scattering [48,54] (Fig. 3c). In this process, two electrons and two holes share the energy of the decayed SPP, so on average, the energy of each carrier is just $\hbar\omega/4$ ("warm" carriers). It is well known that, at low frequencies, EE scattering contribution to the electric resistance (and, therefore, Ohmic loss) is negligibly small. The reason for this is the fact that the total momentum of carriers undergoing EE scattering is conserved: $k_1' + k_2' = k_1 + k_2$. Since as long as the band can be considered parabolic near the Fermi surface, the current density can be found as $J = -e\sum_k v_k = -\frac{e\hbar}{m}\sum_k k$, where $l\ v$ is the electron velocity and $m$ is the electron effective mass. Therefore, the current is conserved and no energy is dissipated via the EE scattering. But for the optical frequencies, the situation is dramatically different. The photon energy is sufficiently large to initiate the Umklapp processes [24,55] in which one of the photoexcited electrons is promoted into the adjacent Brillouin zone so that momentum conservation relation becomes $k_1' + k_2' = k_1 + k_2 + g$, where $g$ is the reciprocal lattice vector. Obviously, the electron velocity and current now change as a result of the EE scattering, hence the process becomes allowed, and the SPP decay. The EE-scattering assisted SPP damping rate can be found as $\gamma_{ee}(\omega) = F_U(\omega)\tau_{ee}^{-1}(\omega)$, where the EE scattering rate is [56]

$$\tau_{ee}^{-1} \approx \frac{\pi}{24}\frac{E_F}{\hbar}\left(\frac{\hbar\omega}{E_F}\right)^2 \qquad (3)$$

and $F_U(\omega)$ is the fraction of the total EE scattering events that are Umklapp processes. This fraction is typically on the order of 0.2—0.5. It follows that the EE-assisted SPP decay becomes prominent at short wavelengths and in most metals for photon energies larger than 2 eV the EE scattering rate in noble metals is $\gamma_{ee} \sim 10^{14} s^{-1}$, i.e., at least as large as the phonon-assisted SPP damping rate as was indeed measured.[48] At the same time, for the photon energies less than 1 eV, the EE-assisted damping is not important. The energy distribution of the "first generation" carriers excited with assistance of the EE scattering is

$$F_{hot,ee} = 2 \times 3(\hbar\omega - E_F)^2 / (\hbar\omega)^3, \qquad (4)$$

where the factor of 2 indicates that 2 hot electrons are excited by each SPP decay event Due to involvement of reciprocal lattice vectors, the angular distribution of the generated electrons is approximately uniform For all three SPP decay mechanisms discussed above, the spatial distribution of non-equilibrium carrier generation simply follows the density of the SPP energy $\mathcal{E}(r)^2$, where $\mathcal{E}(r)$ is the electric field of the SPP.

*Landau damping or surface collision assisted decay*

The fourth and last SPP decay channel (Fig. 3d) is referred to either phenomenologically as surface collision assisted decay or, in a quantum picture, as the Landau damping (LD) [57-60]. Classically, when the electron collides with the surface (or the "wall"), the momentum can be transferred between the electron and the entire metal lattice, in a way similar to what happens when electron collides with a phonon or defect. That relaxes momentum conservation rules and, as first done by Kreibig and Vollmer [61], one can describe it by simply introducing the surface collision rate $\gamma_{sc} \sim v_F / d$, where $d$ is the size of nanoparticle. Quantum mechanically, the absorption is the result of the spatial localization of optical field. Since the field is localized, its Fourier transform contains all the spectral components, some of them higher than $\Delta k = \omega / v_F$, where $v_F$ is Fermi velocity, which for Au and Ag is about $1.4 \times 10^8 cm/s$. These spectral components provide necessary momentum matching which allow decay of LSP/SPP without assistance from the phonons or defects. This process is commonly referred to as LD [10,62,63] and is characterized by the existence of the imaginary part of the wavevector-dependent (nonlocal) dielectric permittivity of the metal described by the Lindhard's formula[64]

$$\varepsilon(\omega, k) = \varepsilon_b + \frac{3\omega_p^2}{k^2 v_F^2}\left[1 - \frac{\omega}{2kv_F} \ln\frac{\omega + kv_F}{\omega - kv_F}\right] \qquad (5)$$

and has the imaginary part for $k > \omega / v_F$. The LSP/SPP decay rate due to the LD is $\gamma_{LD} = 3v_F / 8d_{eff}$ where the effective depth is $d_{eff} = \int_{metal} \mathcal{E}(r)^2 dV / \int_{surface} \mathcal{E}_\perp^2(r) dS$, i.e. is defined by the volume-to-surface ratio of the LSP/SPP mode in the metal, and $\mathcal{E}_\perp(r)$ is the normal to the surface component of the LSP/SPP electric field. Both phenomenological and more exact full quantum treatments provide similar results[57]. For example, for spherical nanoparticles, ne can obtain $\gamma_{LD} = 0.75 v_F / d$ while in the phenomenological treatment it is $v_F / d$.

There are two reasons for LD (surface collisions) being so vital for hot carrier processes. First, the excited hot carriers generate are all located near the interface within a thin layer of thickness $\Delta L = 2\pi / \Delta k = v_F / v$ [41], where $v$ is an optical frequency In other words, $\Delta L$ is the distance covered by the electron over one optical oscillation period. For example, for Au under the 700 nm wavelength excitation, it is only about 3 nm, which is obviously shorter than the mean free path of electron between collisions (typically 10-20 nm). Therefore, one half (the other half travels away from the surface) of the carriers excited via the LD will always end up at the surface, which is definitely not the case for other mechansims. The second reason for LD prominence is that the angular distribution of the carriers excited via the LD is highly nonuniform: $R_{LD}(\theta) \sim 2|\cos^3 \theta|$, The fraction of hot carriers that impinges on the surface at normal incidence is increased by a factor of 4 compared to the uniform distribution (characteristic for interband and e-e scattering assisted processes) and by a factor of 2 compared to the $R_{ph}(\theta) \sim 3\cos^2 \theta / 4 + 1/4$ distribution of the carriers generated by phonon-assisted processes[27]. It should be noted that a simple treatment of LD presented here gives results that are essentially the same as hydrodynamic theory with diffusion terms included, as, for instance in Refs.[65-67].

## Hot-carrier injection at a metal-semiconductor interface.

Let us consider what happens when ½ of all the carriers generated near the surface by the LD as well as a (small) fraction of all the carriers generated by other mechanisms in the bulk of a plasmonic metal impinge onto the metal-semiconductor (or metal-dielectric) interface at an angle $\theta$ (Fig. 4a). The angular distribution of the impinging electrons, $R_{eff}(\theta)$ is typically determined by LD. In the case of a smooth surface, the momentum conservation for the in-plane (lateral) electron wavevector $k_\parallel \approx k_F \sin \theta$ must be maintained. The energy conservation must also be maintained. Therefore, for the electron in the metal having the energy $E$ above the Fermi level, the normal to the interface electron wavevector is $k_{m,z} \approx k_F \cos \theta$, For a semiconductor, the normal to the surface electron wavevector component is $k_{s,z} = \sqrt{2m_L / \hbar^2 (E - \Phi) - (m_L / m_T) k_\parallel^2}$, where $\Phi$ is barrier height, $m_L$ and $m_T$ are the longitudinal and transverse effective electron masses in semiconductors, such as, for example, Si with the band structure consisting of 6 valleys (Fig. 4b) [68]. The maximum angle which still allows the propagation of the electron from metal to semiconductor is $\theta_{max}(E, \Phi) = \sin^{-1} \sqrt{(m_T / m_0)(E - \Phi) / (E + E_F)}$, where $m_m$ is the effective electron mass in the metal, typically close to the free electron mass $m_0$. Using a Au/Si interface as an example, and assuming $E \sim 1\ eV$, $\Phi \sim 0.5\ eV$, $E_F = 5.5\ eV$, and $m_T = 0.2 m_0$, one obtains $\theta_{max} \sim 7°$ corresponding to solid angle of $\Omega_{max} \approx 0.015\pi$, meaning that less than 1% of the incident carriers make it over the barrier. Since the incident carrier energies are typically distributed uniformly in the interval $0 < E < \hbar\omega$, the overall efficiency of the carrier injection is

$$\eta_{ext}(\hbar\omega, \Phi) = \int_0^{\hbar\omega} F_{hot}(E, \hbar\omega) \int_0^{\theta_{max}(E,\Phi)} R_{eff}(\theta) T(\theta, E) \sin\theta d\theta dE \qquad (6)$$

where $F_{hot}(E, \hbar\omega)$ depends on the type of the decay process (for simple estimations can be typically assumed $F_{hot}(E, \hbar\omega) \approx 1/\hbar\omega$) and the angle dependent electron transmission coefficient is

$$T(\theta, E) = 1 - \left( \frac{k_{m,x}/m_0 - k_{s,z}/m_L}{k_{m,z}/m_0 + k_{s,z}/m_L} \right)^2. \tag{7}$$

Note that multi-valley nature of the conduction band in indirect bandgap semiconductor like Si or Ge significantly affects the hot electron injection. Only two longitudinal valleys participate in injection process since the $\mathbf{k}_\parallel \sim 10^8 \, cm^{-1}$ for transverse valleys requires the electrons in the metal to propagate at grazing incidence to surface where the reflection is very high.

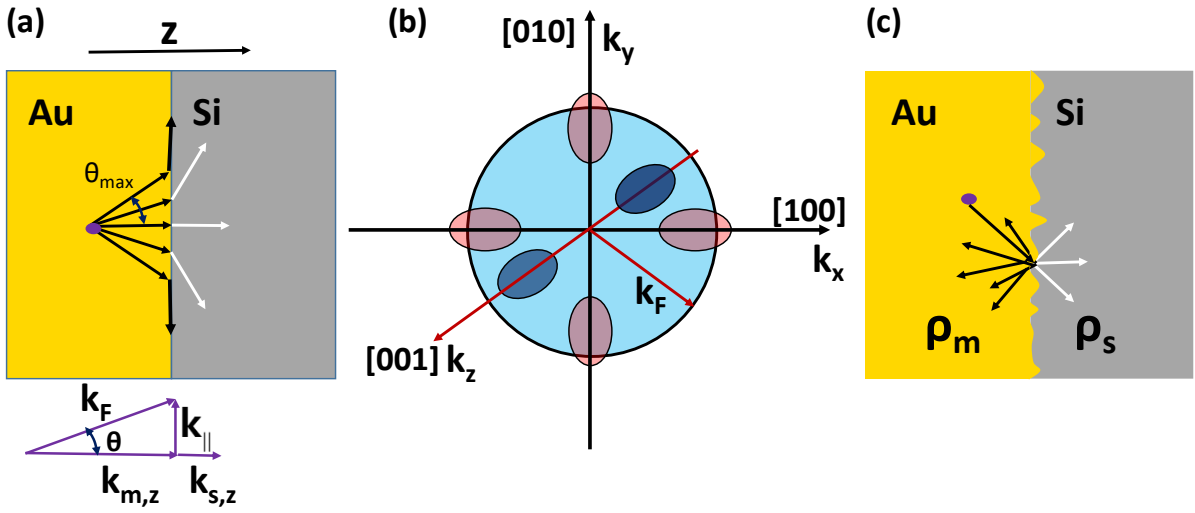

**Fig. 4.** Carrier injection from metal into a multi-valley semiconductor. (a) Carrier injection across a smooth interface into a multi-valley semiconductor (like Si). Injection is possible only if incidence angle less than θ$_{max}$. (b) Conduction band structure of Si. Electrons are injected only into the two valleys along [001] direction. (c) Carrier injection across a rough interface with lifted momentum conservation restrictions.

For small values of the electron incidence angles, the approximation of Eq. 6 leads to the Fowler's formula [69]

$$\eta_{ext}(\hbar\omega, \Phi) \approx \frac{1}{4} R_{eff}(0) T_{eff}(0) \frac{m_T}{m_0} \frac{(\hbar\omega - \Phi)^2}{\hbar\omega E_F}. \tag{8}$$

The results of the extraction efficiency calculations for a smooth Au/Si interface and the SPPs excited with $\hbar\omega = 0.8$ eV ($\lambda = 1500$ nm) show that it never approaches even 1%. However, the experimental data reveal that higher injection efficiencies can be achieved when the momentum conservation is no longer valid due to nanoscale structuring or disorder of the interface on a nanometer level. The injection efficiency of nearly 30% for a Au/GaAs interface was reported[70] and even higher, 45% efficiencies for injection into TiO$_2$ from Au nanoparticles has been measured[71]. Ten-fold increase in photocurrent in the photodetectors with rough Au/Si interface[72] relative to the ones with a smooth interface[73] has been observed (although differences in processing mean that not all of this enhancement can be attributed to the interface roughens).

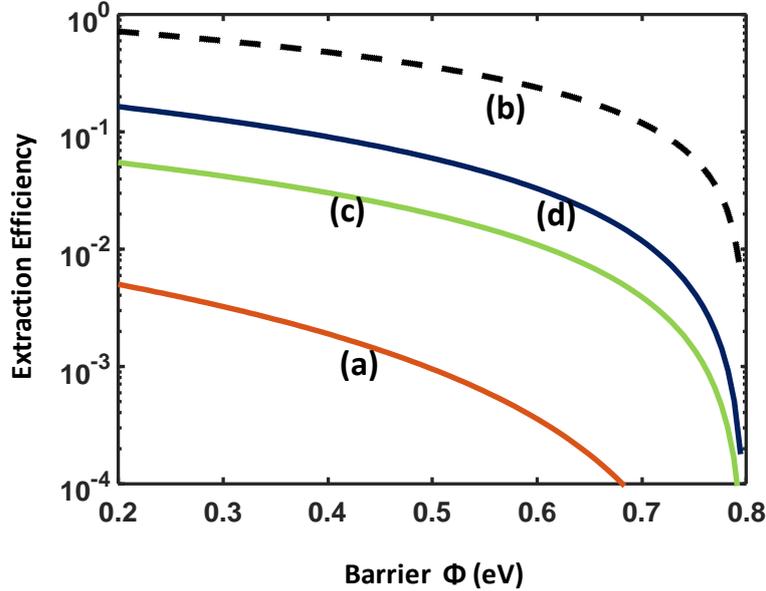

**Fig. 5.** The dependence of the extraction efficiency of the hot carriers at an Au/Si interface on barrier height. (a) smooth interface (Eq. 8), (b) rough interface with complete extraction of all the above-the-barrier carriers (Eq. 9), (c) rough interface with the momentum conservation rules relaxed but the transitions allowed only into 2 longitudinal valleys of Si (Eq. 10), and (d) rough interface with the momentum conservation rules fully lifted and the transitions into 4 transverse valleys of Si allowed (Eq. 11). The excitation wavelength is λ=1500 nm.

These results can be explained assuming that nearly all the hot carriers with energies higher than barrier Φ can be extracted in the semiconductor [70]

$$\eta_{ext}(\hbar\omega, \Phi) = \hbar\omega/\Phi - 1 \tag{9}$$

This assumption, however neglects the possibility of electron backscattering into the metal at the interface. More rigorous approach considers explicit description of the electron scattering on interface roughness demonstrating the enhancement of extraction efficiency by a factor of a few[74]. The model is only applicable to a relatively small roughness and neglects the backscattering as well.

To estimate the electron extraction efficiency more rigorously, the theory developed for the seemingly different task of light trapping in the dielectric with roughened surface can be applied[75]. Since for a rough surface the momentum conservation is no longer valid, according to the Fermi's golden rule, the rate of scattering in a given direction depends only on the density of states (Fig. 5c). If the density of states in the metal and semiconductor are $\rho_m$ and $\rho_s$ respectively, the extraction efficiency can be derived to be[27]

$$\eta_{ext,\max}(\hbar\omega,\Phi) = \int_0^{\hbar\omega} f_1(E,\hbar\omega) \frac{\rho_s(E)}{\rho_s(E)+\rho_m(E)} dE = \frac{1}{\hbar\omega}\int_0^{\hbar\omega} \frac{(m_{DOS}/m_0)^{3/2}(E-\Phi)^{1/2}/E_F^{1/2}}{(m_{DOS}/m_0)^{3/2}(E-\Phi)^{1/2}/E_F^{1/2}+1} dE$$
(10)

where $m_{DOS}$ is the density-of-states effective mass of semiconductor[76]. For Si, if one assumes that the injection takes place in only 2 valleys along [001] direction (Fig. 4b), $m_{DOS} = 2^{2/3}(m_L m_T^2) = 0.52 m_0$, resulting in injection efficiency shown in Fig. 5c. If, on the other hand the spatial spectrum of roughness contains high spatial frequencies on the scale of $1/k_F \sim 1\text{Å}$, then all 6 valleys can receive the injected hot carriers and $m_{DOS} = 6^{2/3}(m_L m_T^2) = 1.08 m_0$. In the case of small extraction probability,

$$\eta_{ext}(\hbar\omega,\Phi) \approx \frac{2}{3}\left(\frac{m_{DOS}}{m_0}\right)^{3/2} \frac{(\hbar\omega-\Phi)^{3/2}}{\hbar\omega E_F^{1/2}} .$$
(11)

The extraction efficiency (Fig. 5d) estimated with Eq. (11) is not as high as expected from Eq. (9) yet it is higher than the extraction efficiency for smooth surface described by Eq. (8). Assuming the barrier height of 0.5 eV, one can see that the enhancement by a factor of 20 to 60 can be possible leading to extraction efficiencies approaching 10%, A significant factor in the enhancement is the fact that the DOS mass is significantly larger than the transverse mass in Si, but even for the semiconductors with relatively large isotropic mass, such as II-VI materials, the enhancement should be significant.

The key point to be taken from here is that ultimately it is density of states that determines the injection efficiency, no matter whether interface is smooth or not. Choosing semiconductor with larger effective mass (for TiO$_2$ it is comparable to m$_0$[77]) and metal with a relatively low Fermi energy (TiN with low density of states at the Fermi level[78] comes to mind) can be highly beneficial for injection efficiency.

## Hot-electron chemistry

In addition to their application to photodetection, hot carriers have been recently extensively studied as a means to control chemical transformations in photochemical and photocatalytic settings[79-83]. In this context, different mechanisms need to be considered to understand the role of hot carriers and differentiate it from other processes present at the same time in the nanoplasmonic environment. These include: (i) molecular adsorption on a plasmonic material resulting in the modification of the energy levels of the molecules and therefore changing its reactivity, which can additionally be accompanied by the local temperature increase of the nanostructure upon illumination. The modification of the energy levels upon adsorption is a complex process, which is also influenced by the electrons in a metal near the Fermi level and, as such, not much modified by the optical excitation which changes the density of electrons near the Fermi level a little; (ii) simple increase of light-absorption by an adsorbed molecule in the vicinity of a plasmonic nanoparticle due to the field enhancement effect, which results in a conventional photochemical processes induced by light[84]; (iii) a charge (electron or hole) transfer to the adsorbed molecule; followed by (iv) product desorption. During the process (iii) in particular, a transient charged state of an adsorbate is formed with its own energy states (which are obviously different from the non-charged state) and which have different chemical reactivity. This carrier injection can occur either through the LSP/SPP decay inside a plasmonic nanostructure with subsequent injection through the interface (Fig. 6a) to the molecule (similar to the injection into a semiconductor discussed above) or through hybridised surface states) or the presence of the adsorbate on the metal surface provides additional channel of the LSP/SPP decay (similar to the defect scattering process discussed above) with the electron appearing directly in the electron accepting orbitals[85] (LUMO). The latter process (Fig. 6b,d), sometimes referred to as 'chemical interface damping (CID) [86,87] is favourable from the point of view of preserving the hot-carrier energy and the efficiency of it can be comparable to the injection[85,88,89].

The plasmon-induced processes allow controlling speed, efficiency, and activation barriers of chemical transformations (Fig. 6c-h), therefore there arises a possibility of controlling selectivity and final products of the reactions. While in many cases all four contributions (i)-(iv) are important for the chemical reaction control and should all be carefully considered, the electron transfer is required in the case of oxidative/reduction (redox) reactions, and these are particularly affected by the engineered hot-electron injection. Injection of hot electrons in a surrounding solvent (even vacuum) can be achieved. During ballistic propagation in the surrounding medium, electron energy will be preserved, therefore, adsorption of molecules is not the necessary condition for inducing chemical reactions as long as the ejected electron will interact with a reactant before losing its energy. This can be used for reaction involving surrounding medium itself, such as, for example, water splitting or reactive oxygen species generation[90]. In most cases, however, the adsorption significantly facilitates the interaction between hot-electrons and reactant molecules. The adsorption of a reactant and desorption of the reaction product, needed to free site for the next cycle of the reaction, themselves depend on the local heating and molecular hybridisation with a nanostructure.

The ideal design of a plasmonic catalyst should provide (A) efficient light absorption in a plasmonic material, which is relatively easy to design near a plasmonic resonance; (B) the efficient generation of the hot-carriers (preferably through the Landau-damping as was discussed above); (C) an easy route for the hot carriers to leave a plasmonic metal and interact with the species adsorbed on a surface; and therefore, (D) a surface with an appropriate surface energy for efficient adsorption. Conditions A–C can be optimised for a required illumination wavelength, which determines the energy distribution of hot carriers by the choice of a plasmonic metal, as well as the size and topology of the nanoparticles. Condition D is difficult to satisfy using plasmonic metals for which the electronic d-bands are far in energy from the Fermi level, where the bonding and anti-bonding states of an adsorbate are fully occupied, and, therefore, no metal–adsorbate interactions are likely to occur. At the same time, in typical catalytic metals, a d-band is close to the Fermi energy, promoting effective adsorption, but for the same reason these metals are poor plasmonic metals in the visible and near-infrared spectral ranges. Therefore, a combination of plasmonic and catalytic nanoparticles in a hetero-nanoparticle may be needed to satisfy all conditions A–D. Moreover, such combination may also improve condition B by creating an additional electromagnetic field enhancement near the nanoparticle junctions, where hot carriers are generated more efficiently. Small (few nm-size) catalytic nanoparticles on a surface of larger (few 10s of nm) plasmonic nanoparticles also provide good conditions for effective hot-carrier extraction by supplying an additional momentum to hot-electrons. In this way, plasmonically-derived hot carriers are most efficiently generated and extracted at the locations where the adsorbates are positioned[91]. The other considerations for the photocatalytic efficiency are to provide large surface area, so that nanoparticles are preferable to a planar surface, and to avoid semiconductor materials as, for example, a substrate that could trap the excited hot carriers. As opposed to metal–semiconductor heterostructures exhibiting interfacial Schottky barriers through which hot carriers have to tunnel, no Schottky barrier is present in bi-metallic nanostructures, leading to the availability of more hot carriers for catalytic processes.

The figure of merit ($FOM_{PC}$) for hot-electron-induced photocatalysis can be introduced considering various processes involved in the promoting chemical reaction. Experimentally, however, care should be taken comparing different performances, which may be limited by other than hot-carrier processes related factors, such as reactant diffusion and adsorption and product desorption rates, as well as influenced by local temperature increase due to light absorption in plasmonic nanoparticles. However, a FOM of the photocatalytic action of the hot carriers produced in plasmonic hetero-nanoparticles can be evaluated through the hot-electron injection rate determined by the light absorption rate $\gamma_{abs}$ in plasmonic nanoparticles, the efficiency of the hot carrier generation $\eta_{hc}$, which depends on the illumination wavelength (through the LSP resonant conditions), and the efficiency of the hot-electron transfer to the adsorbate $\eta_{extr}$: $FOM_{PC} \sim \gamma_{abs}\eta_{hc}\eta_{extr}$. In turn, the light absorption rate depends on the rate of the incident photons of the illuminating light (peak intensity, $I$) and the absorption cross-section of the nanoparticle, $\sigma_{abs}$. Therefore, for the given conditions, the photocatalytic efficiency of the nanoparticles can be optimised by maximising the expression $FOM_{PC} \sim I \sigma_{abs}\eta_{hc}\eta_{extr}(r_0)$, where $r_0$ indicates that an adsorbed molecule and hot-carrier extraction should be in the same location. Generally, for the optimal designs, the presence of catalytic nanoparticles (few nm size) on a surface of plasmonic nanoparticles (20-50 nm) barely influence the absorption which is dominated by a plasmonic resonance. At the same time these plasmonic sizes are advantageous for optimising the Landau damping generation of hot-carriers, while nanometric catalytic metal facilitate extraction from Au to Pt-molecular hybrids. This also clearly shows the advantage of pulsed illumination as opposed to CW illumination, with the pulse duration preferably shorter than LSP decay. For complex shapes of plasmonic nanostructures, such as nanoshells, nanorods, nanopiramids, nanostars or bow-tie type antennas, the dependence of the LSP resonance position, illumination polarisation (if any) are also important since the hot-carrier generation is most efficient at the field enhancement hot-spots and indeed the distance to the surface where extraction take place can be controlled.

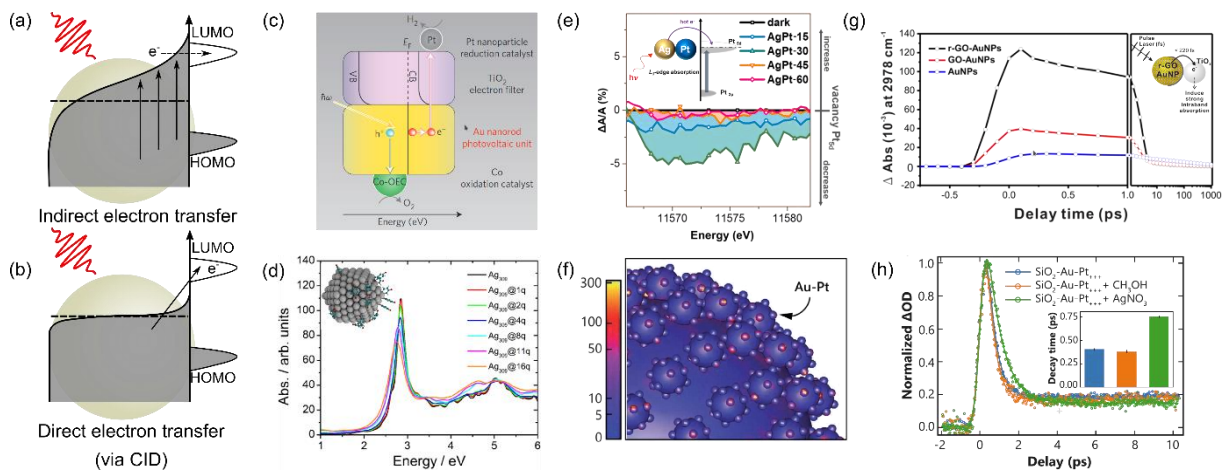

**Fig. 6.** Hot electrons in chemical processes. (a,b) Two mechanisms of hot carrier excitations in molecular adsorbates: (a) Hot electron injection and (b) Direct electron transfer (CID) (c) Design of a metal/semiconductor device for water splitting: gold nanorod is used as a source of plasmonic hot electrons, TiO$_2$ as a hot electron filter, smaller Co and Pt nanoparticles act as catalysts for oxygen and hydrogen evolution[80]. (d) Atomistic calculation of broadening of plasmonic resonance of a silver cluster due to chemical interface damping (CID), induced by coupling with quinine molecules: broader peaks are obtained with higher quinine coverage[87]. (e) XANES spectra, illustrating plasmonic hot-carrier transfer in photocatalytic bimetallic Ag/Pt plasmonic alloys: optical illumination leads to a decrease in Pt$_{5d}$ vacancy concentration affecting the reaction pathway[93]. (f) Design of plasmonic Au/Pt hetero-nanoparticles for optimized plasmocatalytic response: photocatalytic units comprised of larger plasmonic gold nanoparticles and smaller Pt nanoparticles are deposited on surfaces of large supporting silica particles[94]. (g-h) Transient absorption spectra for monitoring hot-carrier dynamics during chemical reactions: (g) Au/reduced graphene oxide(r-GO)/TiO$_2$ photocatalyst-transient absorption signal originating in TiO$_2$ from injection of plasmonic hot electrons from gold nanoparticles depends on the intermediate layer thickness[91]; (h) Gold/platinum hetero-nanoparticles-decay of hot-carrier population in gold depends on the presence of hot-electron scavenger AgNO$_3$ but not hot-hole scavenger CH$_3$OH[92].

## Hot-electrons and nonlinear optical effects

Even a single LSP excitation results in the modification of the electron gas distribution and associated changes of the metal permittivity and refractive index. Under strong femtosecond photoexcitation, as outlined above, multiple LSP excitation may be achieved, eventually leading to a highly nonequilibrium hot electron population in the conduction band of a metal. This hot carrier population, inherently nonequilibrium for the first few hundred of femtoseconds (as outlined in the previous sections), subsequently decays at picosecond timescale through emission of phonons, and can be experimentally monitored using time-resolved pump-probe measurements[95-100].

The optical nonlinearity arises in this context from a strong dependence of the dielectric function of the metal on the distribution (for non-thermalised hot carriers) or temperature (after thermalisation) of hot electrons. Indeed, the dependence of the dielectric constant on the intensity of light produces effective cubic Kerr-type nonlinearity, that allows us to envision nanoplasmonic systems operating as optical switches for intensity, phase or polarisation. The dielectric function of most metals is most sensitive to the excitation of hot carriers in the visible range, close to the offset of direct interband transitions from the d-band to the vicinity of the Fermi level (e.g., around 2.35 eV for gold). Experimentally, the related "smearing" of the Fermi distribution produces a photobleaching signal above the threshold due to band filling as well as photoinduced absorption below the threshold [97,101]. Theoretically, this can be treated in a straightforward way employing the Fermi's golden rule for direct interband transitions [36,102] or a full density functional approach. The variation of dielectric constant of prototypical plasmonic metal (gold) under strong excitation is shown in Fig. 7d. In the near-infrared, on the other hand, the excitation of hot carriers within the conduction band (intraband excitation) mainly contributes to the increase of the bulk damping (that appears in the Drude expression of optical conductivity of metal) due to the electron temperature dependent Umklapp electron-electron scattering[25,47]. The latter process is usually weaker and often not observed from smooth metal surfaces outside the surface plasmon excitation condition, however the local fields enhancement in plasmonic systems allows to enhance this effect[103]. Furthermore, because Fermi level in metals lays close to the boundary of the Brillouin zone, hence the conduction band is non-parabolic and hot carriers have different effective mass from the equilibrium electrons. This affects the plasma frequency and, through it, the dielectric constant and is the main source for hot-electron driven ultrafast manipulation of epsilon-near-zero materials[104,105]. Furthermore, low free carrier concentrations in prototypical ENZ materials allow for much faster hot carrier relaxation and therefore higher switching speeds[104,106].

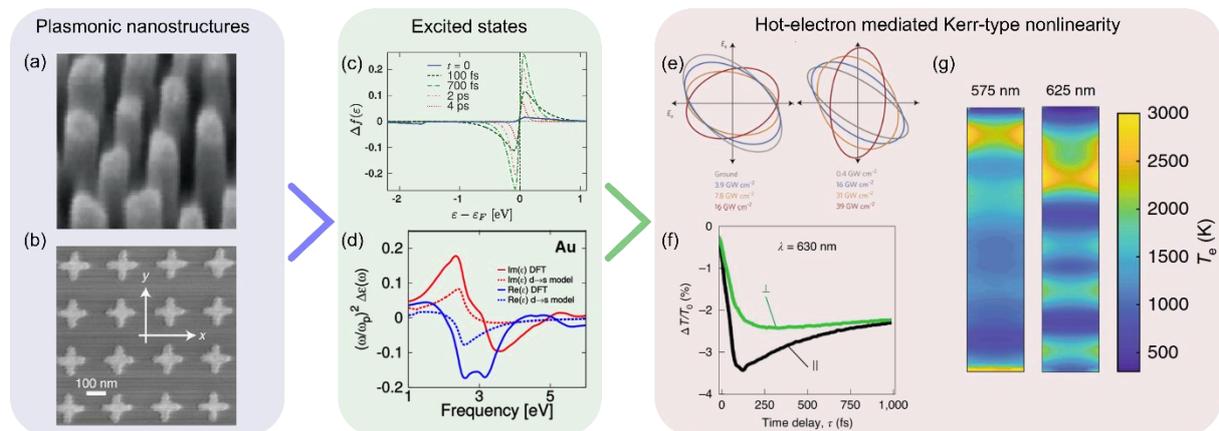

**Fig. 7.** From plasmonic metasurfaces to ultrafast nonlinearity. (a,b) Plasmonic metamaterials/metasurfaces used to active control of light: (a) nanorods[103,107], (b) nanocrosses[108]. (c) Distributions of hot electron energies near the Fermi level calculated from the first principles at different times after photoexcitation of a gold nanoparticles[109]. (d) Changes of the real and imaginary parts of the permittivity of gold calculated using different models induced by an electron temperature of 5000 K [102]. (e) Optical control of polarization of light using hot-electron induced nonlinearity in a nanorod metamaterial near the ENZ wavelength[103]: (left) pump-induced and (right) self-induced polarisation changes for different excitation powers. (f) Induced anisotropy in pasmonic nanocrosses, observed through differential transmission for aligned (black) and crossed (green) polarisations of control/probe beams[108]. (g) Local, wavelength-dependent hot-electron distributions inside gold nanorods in the metamaterial, which determine optical response dynamics[107].

The efficiency and speed of hot-carrier-driven nonlinear nanoplasmonic devices is linked to the dynamics of the hot carrier decay, it is, therefore, important to discuss this process. Since the electron thermalization time, $\tau_E$, is at least an order of magnitude smaller than hot electron cooling rate $\tau_{EL}$, different stages of the dynamics (Fig. 2a-e) can be separated with ease. Strong excitation producing dense hot carrier population in the conduction band facilitates "collective" thermodynamic description of the evolution based on the energy exchange between hot electrons and lattice modes (phonons). This leads to a formulation of a simple phenomenological model known as the two-temperature Model (TTM) [95-97,102,110-113]. Within the scope of the TTM only the heat exchange between electrons and lattice, described by their corresponding temperatures ($T_e$ and $T_l$, respectively) can be considered coupled through the effective electron-phonon coupling constant $G$. The model further takes into consideration the nonlinear temperature dependence of the hot electron specific heat, derived from the free electron model, and the dependence of the lattice heat capacity on $T_l$ given by the Debye theory. However, for most noble metals and experimental conditions, the rise in electron temperature does not exceed 0.1 $T_F$ (Fermi temperature, around $6.4 \times 10^4$ K for gold for example [68]), therefore, a simplified expression for $C_e = \gamma T_e$ may be used. These physical considerations lead to the following set of equations for electron and lattice temperatures:

$$\gamma T_e \frac{dT_e}{dt} = \nabla(D_e(T_e, T_l)\nabla T_e) + G(T_l - T_e) + S(t)$$

$$C_l(T_l)\frac{dT_l}{dt} = G(T_e - T_l),$$

where the two additional terms in the first equation describe heat diffusion due to electron thermal conductivity $D_e$, and a source term, describing heating by a strong and short laser pulse. Most of the present works rely on TMM parametrization for hot electron heat capacity and electron-phonon coupling constants derived from experiment or microscopic estimates of the free electron model [112], an approach to use the parametrisation derived from more rigorous ab-initio calculations have been demonstrated recently [102].

The hot electron-induced optical transients have been extensively used not only as means for studying the hot electron dynamics in metals but as a means of active control of optical Kerr-type nonlinearities in the plasmonic systems designed for potential practical applications. A great number of experimental geometries have been developed for the purpose of all-optical switching applications, including, among others, localized surface plasmon based systems[108], surface plasmon polaritons[114,115], uniaxial plasmonic composites, made of nanowires[103,107,116]. The geometry of the nanostructures provides additional degrees of freedom for ultrafast manipulation of optical response. Geometries like nanocrosses or aligned plasmonic nanorods (Fig. 7a-b) allow for control of optical polarisation on picosecond time scale (Fig. 7e-f) due to low symmetry of the corresponding structure. More remarkably, as will be discussed below, introducing spatial degrees of freedom in optically thick nanoplasmonic composite (Fig. 7g) allows manipulation of the optical polarisation at even sub-picosecond timescale surpassing the limits imposed by the time constants of hot electron decay alone. Recently, plasmonic nanostructures coupled to epsilon-near-zero materials, such as indium tin oxide, aluminium zinc oxide and others, are widely explored for efficient optical switching applications[106]. In particular, for the realisation of the time-varying response and time varying metamaterials, the nonparabolic conduction bands in such oxide materials provides short electron relaxation times and very large nonlinear optical changes of the refractive index.

Evolution of the hot electron population in space, occurring when the dimensions of the structure exceed the penetration depth of the optical field has been also a subject of extensive research in metals and plasmonic nanostructures. In simple experimental geometries, such as metal films "time-of-flight" experiments were used to determine the electronic thermal conductivity $D_e(T_e, T_l)$ that appears in the refinement of the TTM[117,118] as well as to demonstrate superdiffusive hot electron transport on distances comparable to the electron mean free path (40-50 nm [117]). While such transport phenomena have no charge current associated with them and only cause heat dissipation (timescale for dynamic screening in free electron gas is $\tau_{scr} \approx \omega_p^{-1}$, and is in femtosecond domain[119]), ballistic hot electrons can nonetheless carry pure spin current if the noble metal film is interfaced with the ferromagnetic metal[120,121], bordering nonlinear plasmonics with the field of spintronics. Hot electron transport phenomenta can be beneficial in controlling the Kerr-type nonlinearity of plasmonic composites. For example, transport across the ensemble of long gold nanowires[107], allows to manipulate temporal response through variable coupling between the optical modes and hot electron spatial distribution (the examples of such distributions are shown in Fig. 7g). Buried features in metallic composites[122] can be seen through opaque layers due to modifications of transient response induced by hot carrier transport .

**Beyond the Two-Temperature model**

While the two-temperature model is only capable of treating the hot carrier decay phenomenologically it has been extremely successful in describing various light-induced phenomena in photoexcited metals and plasmonic nanostructures owing to its remarkable simplicity. For the same reason it is very easy to extend the TTM to account for additional interactions or ensembles. Modifications involving nonzero phonon thermal conductivity[123], temperature-dependent electron-phonon coupling[96], spatially varying material parameters[122] and even interaction with nonequilibrium hot electron ensemble[98,124] have been successfully demonstrated. For ferromagnetic metals the TTM model can be extended to include the third temperature attributed to the spin population[125].

While the TTM successfully describes the cooling of the thermalized hot electron population within $\pm k_B T_e$ (see Fig. 2c) around the Fermi level it cannot account for the initial nonequilibrium stage of the hot carrier decay in metals. While this population is short-lived understanding it is extremely important as it contains the electrons(holes) with large excess energies above(below) the Fermi level which as mentioned in the previous sections are contributing to the injection into the semiconductor or can drive chemical reactions in the adsorbed species. The dynamics of this non-thermal hot electron ensemble could either accounted for through a simple rate-equation extension of the TTM as proposed by Sun[98] (currently often referred to as "Three-temperature model") or, more accurately, using the Boltzmann transport equation:

$$\frac{\partial f(\mathbf{r},\mathbf{k},t)}{\partial t} = \mathbf{r}\frac{\partial f(\mathbf{r},\mathbf{k},t)}{\partial \mathbf{r}} + \mathbf{k}\frac{\partial f(\mathbf{r},\mathbf{k},t)}{\partial \mathbf{k}} = \Gamma_E[f](\mathbf{r},\mathbf{k}) + \Gamma_{EL}[f](\mathbf{r},\mathbf{k})$$

where $\Gamma_E/\Gamma_{EL}$ are collision integrals for electron-electron and electron-phonon scattering respectively. These collision integrals are extremely computationally demanding and for practical applications the electron-phonon scattering is almost exclusively treated in a relaxation anzats[98,126] (even though more accurate phonon collision integrals have been

used[127,128]), while electron-electron scattering is described within the scope of the Fermi liquid theory[98,126]. More accurately, the collision integrals can be parametrised with ab-initio-derived scalar constants[109]. This approach produces the dynamics of instantaneous energy distribution of hot carriers (Fig. 7c), and is useful for studies of hot-electron injection into semiconductors[129]. The Boltzmann transport approach outlined here is currently limited to metal nanoparticles and thin films where the spatial variation of the hot electron distribution can be ignored. This approach has been successfully employed to investigate the nonequilibrium hot carrier dynamics in various metallic thin films and particles including silver, gold[98-100], copper[130], chromium[113] among others. It was demonstrated, that electron thermalisation time can be controlled in plasmonic metals with the strength of the excitation [109], as the lifetime of excited states is inversely proportional to its energy with respect to the Fermi level. Furthermore, in nanoparticles smaller than 20 nm in diameter, dynamic screening in the electron spill-out region at the surface, was shown to strongly affect the hot electron thermalisation[131-134].

## Conclusions and outlook

Plasmonic nanostructures harness the properties of free-electron oscillations in metals and highly doped semiconductors in order to provide strong electromagnetic field confinement and enhancement near the interfaces. This enhancement in turn assists in driving the electron gas out of equilibrium and enables a number of transient optical phenomena occurring on various timescales. As we have shown above, the nonequilibrium dynamics of hot carriers excitation and decay are determined by the quantum nature of the processes as well as material properties and driven by the coupling between electronic, phonon and structural degrees of freedom, upon the optical irradiation.

At picosecond timescales, the nonlinear response of plasmonic structures is defined by optically-induced heating and subsequent cooling of the free-electron gas. The speed of the latter process and resulting Kerr-type nonlinearity, is defined by the relaxation of the electron temperature and is limited by the electron-phonon scattering and interaction with surroundings. On a sub-picosecond timescale after the excitation, the electron distribution is non-thermal and relaxes much faster via electron-electron scattering. This opens a pathway to non-equilibrium plasmonics with applications in all-optical control of optical signals and plasmonic photochemistry among others. At femtosecond and sub-femtosecond domains, short electron bursts produced by ultrafast photoemission from plasmonic nanostructures can be used as electronic gates for ultrafast electronics. Further bridging the gap between attosecond physics and plasmonics, localised high-harmonic generation in gases assisted by plasmonic field localization has been considered.

The future development of and requirements on nonequilibrium plasmonics can be envisaged in both the implementation of particular applications of tuneable and switchable electromagnetic fields, control of chemical processes, as well as a tool for studying new effects in molecular, atom and material physics. A persistent challenge for implementation of ultrafast plasmonic nonlinearities in various metamaterial and metasurface designs still evolves around enhancing the nonlinear response in order to reduce the required light intensities and control (reduction) of the switching times. Both tasks can be addressed at both material level and nanostructuring as well exploitation of subtle properties of nanostructured media and their hybridisation with molecular or atomic species. A search for new plasmonic materials with a tailored free-carrier concentration and/or non-parabolicity of the conduction band is essential for enhancing the nonlinearity. Influencing electron-electron and electron-phonon scattering rates provide an access to tailored dynamics of hot-electrons, thus providing the control of the nonlinearity response time. Even for the same material, nanostructures with anisotropic electron diffusion can be used to reduce the signal switching time. Hetero-nanostructures providing additional channels for electron relaxation in adjacent materials are also efficient in controlling the nonlinearity temporal response. Similarly, an appropriately chosen environment which provides a channel for hot-carrier sink can be exploited. Ultrafast magneto-plasmonic, nano-photochemistry, ultrafast quantum optics are also important strands of applications exploiting dynamical properties of hot-carriers.

For established plasmonic materials, such as gold, silver and copper, the progress in fabrication and development of high quality single crystalline and ultrathin films is essential for reducing absorption losses. The availability of single crystalline ultrathin films approaching two-dimensional limit (not possible with traditional sputtering), with atomically smooth interface quality will influence the carrier nonequilibrium dynamics and also provide enhanced nonlinear response due to electron gas confinement and influencing scattering processes [10]. Ultrasmooth plasmonic surfaces will also allow exploitation of full potential of hybridisation with 2D materials, such as TMDCs. Using 2D electron gas and associated graphene plasmons may allow reduction of the all-optical switching speeds and energies in the IR spectral range.

Related to the improved quality of materials and the trend toward hetero-nanostructures and molecular interactions is the requirement on improved theoretical treatment of the non-equilibrium processes and especially important in the case of hybrid molecular-plasmonic structures, where treatment of realistic processes at the interfaces is required, including non-equilibrium electron gas, the enhanced electromagnetic field and electron transfer between the molecular and plasmonic components.

Ultrafast plasmonics has provided numerous unique functionalities and advantages for design and application of nanostructures, metasurfaces and metamaterials for controlling hot-electron dynamics and the interactions of the nonequilibrium carriers with surroundings. Initially used as a means for fast reconfigurability and active functionalities employing nonlinear response, nonequilibrium electron gas in plasmonic nanostructures has been used in development of new opportunities for meta-devices applications in ultrafast gating and imaging, Kerr mode-locking, optical frequency combs, optical parametric generators and oscillators, polarisation control, phase conjugation, adiabatic frequency shifts, anywhere where strong nonlinearity, ultrafast response and engineered amplitude, phase and polarisation of light is required. Non-equilibrium plasmonics has also opened new application areas for metasurfaces to be exploited in photochemistry and photocatalysis where electromagnetic mode engineering and hot-electrons are a powerful tool for influencing the chemical reactivity. The development of our abilities to further control hot-carrier dynamics through the choice of materials, nanostructuring, and environment will further facilitate adoption of the ultrafast linear and nonlinear plasmonic devices, fast nanophotonic components, nanochemistry applications and will be a basis for development of new applications.

**Acknowledgements.** This work was supported in part by the EPSRC projects RPLAS (EP/M013812/1) and CPLAS (EP/W017075/1).